\documentclass[10pt,a4paper]{article}   
\usepackage{latexsym}
\usepackage{amssymb,amsfonts}
\usepackage[dvips]{graphicx}
\usepackage{cite}
\usepackage{bm}        %% bold mathematics
\usepackage{amsmath}   %% non gradito a latex2html!
\catcode`\@=11
\@addtoreset{equation}{section}

\def\maketitle{\thispagestyle{empty}\setcounter{page}0\newpage
                \renewcommand{\thefootnote}{\arabic{footnote}}
                  \setcounter{footnote}0}
\renewcommand{\thanks}[1]{\renewcommand{\thefootnote}{\fnsymbol{footnote}}
               \footnote{#1}\renewcommand{\thefootnote}{\arabic{footnote}}}

\renewcommand{\title}[1]{\begin{center}\Large\bf #1\end{center}\rm\par\bigskip}
\renewcommand{\author}[1]{\begin{center}\Large #1\end{center}}
\newcommand{\address}[1]{\begin{center}\large #1\end{center}}
\newcommand{\pacs}[1]{\smallskip\noindent{\sl PACS numbers:
                       \hspace{0.3cm}#1}\par\bigskip\rm}
\def\babs{\hrule\par\begin{description}\item{Abstract: }\it} 
\def\eabs{\par\end{description}\hrule\par\medskip\rm}

\def\beq{\begin{eqnarray}}    %%%  begequation/eqnarray
\def\be{\begin{eqnarray}}
\def\eeq{\end{eqnarray}}      %%%  endequation/eqnarray
\def\ee{\end{eqnarrayn}}
\def\at{\left(}               %%%  open (
\def\aq{\left[}               %%%  open [
\def\ct{\right)}              %%%  close )
\def\cq{\right]}              %%%  close ]
\renewcommand{\Im}{\mathop{\rm Im}\nolimits} 

\def\be{\begin{equation}}
\def\ee{\end{equation}}
\def\bea{\begin{eqnarray}}
\def\eea{\end{eqnarray}}

\renewcommand{\title}[1]{\begin{center}\Large\bf #1\end{center}\rm\par\bigskip}
\renewcommand{\author}[1]{\begin{center}\Large #1\end{center}}

\begin{document}

\title{On the Hawking radiation as tunneling for a class of dynamical
  black holes}  
\author{R.~Di Criscienzo$^a$,
  M.~Nadalini$^a$\thanks{nadalini@science.unitn.it},  
L.~Vanzo$^a$\thanks{vanzo@science.unitn.it},
  S.~Zerbini$^a$\thanks{zerbini@science.unitn.it}, G.~Zoccatelli$^a$}  
\address{
$^a$Dipartimento di Fisica, Universit\`a di Trento and INFN,  
Gruppo Collegato di Trento, Italia}

\begin{abstract}
The instability against emission of massless particles by the trapping 
horizon of an evolving black hole is analyzed with the use of the
Hamilton-Jacobi method. The method automatically selects one special
expression for the surface gravity of a changing horizon. Indeed, the
strength of the horizon singularity turns out to be governed by the
surface gravity as was defined a decade ago by Hayward using Kodama's
theory of spherically symmetric gravitational fields. The theory also
applies to point masses embedded in an expanding universe, were the
surface gravity is still related to Kodama-Hayward theory. As a bonus
of the tunneling method, we gain the insight that the surface
gravity still defines a temperature parameter as long as the evolution
is sufficiently slow that the black hole pass through a sequence of
quasi-equilibrium states. 
\end{abstract}

\pacs{04.70.Bw, 04.70.Dy}
%\keywords{Black holes, tunnelin method, Hawking radiation}

\section{Introduction}

With the exception of few lower dimensional models where dynamical
computations can be done, the semi-classical theory of black hole
radiation and evaporation has perhaps reached a satisfactory state of
development only for the case of stationary black holes. It includes
now a vast range of topics, ranging from the various derivations of the
quantum Hawking's effect to the efforts aiming to a statistical
interpretation of the area law and the associated thermodynamical
description. One of the surprising aspects of these results is that the
radiation caused by 
the changing metric of the collapsing star approaches a steady
outgoing flow for large times, implying a drastic violation of energy
conservation if one neglects the back reaction of the quantum radiation
on the causal structure of spacetime. But the back reaction problem
has not been solved yet in a satisfactory way. As pointed out by
Fredenhagen and Haag long ago \cite{Fredenhagen:1989kr}, if the back
reaction is taken into account by letting the mass of the black hole
to change with time, then the radiation will originate from the
surface of the black hole at all times after its formation and ``it
will no longer be precisely calculable from a scaling
limit'' \cite{Fredenhagen:1989kr} (the term originates from the link
between the Hawking's radiation and the short distance behaviour on
the horizon of the two-point function of the quantum field). Thus one
is concerned to show, in the first place, that at least some sort of
instability really occurs near the horizon of the changing black
hole. This question is non trivial since a changing horizon is not
even a null hypersurface, although it is still one of infinite
red shift. In this vein we shall analyze this question for a class of
dynamical black hole solutions that was inspired by problems not
directly related to black hole physics, although these were
subsequently reconsidered in the light of the black hole back reaction 
problem in the early Eighties. The metrics we shall consider are the
Vaidya radiating metric \cite{vaidya}, as revisited by
J.~Bardeen \cite{bardeen} and J.~York \cite{york}, together with what
really is a fake dynamical black holes, the McVittie solution
representing, in author's mind, a point mass in
cosmology \cite{mcv}. And the strategy we shall use 
is a variant of a by now well known method due to F.~Wilczek and
M.~Parikh \cite{parikh} according to which the process of Hawking
radiation is akin to a tunneling effect through the horizon. The method
was refined and extended to more general cases in \cite{svmp} and
others papers as well \cite{others}. For criticism and counter
criticism see also \cite{anm}. It must be kept in mind that
these solutions do not correspond precisely to the standard notion of
a black hole, but they do have horizons (apparent and/or trapping) to
which the familiar black hole theorems seems to
apply \cite{Hayward:1993wb}.  

The variant referred to above is the Hamilton-Jacobi
method introduced in \cite{angh}, so called after the comparison
analysis with the Parikh-Wilczek method done by B.~Kerner and
R.~Mann \cite{mann}.  The tunneling
method provides not only new physical insight to an understanding of
the black hole radiation, but is also a powerful way to
compute the surface gravity for a vast range of solutions. 
As a matter of facts, in the past decade some different definitions of 
the surface gravity of an evolving black hole were proposed
\cite{Hayward:1993wb}, \cite{Fodor:1996rf}, until 
one which met with all requirements was introduced in
\cite{Hayward:1997jp} using a key ingredient invented by H.~Kodama
\cite{Kodama:1979vn} (more on this later). The HJ method can also be
applied to the more elaborate theory of isolated horizons of Ashtekhar
and co-workers (see \cite{isohor} and references
therein. The literature is quite extensive). Results in this
direction appeared recently in \cite{gaowu}, where the Parikh-Wilczek
and HJ methods are compared and showed to agree. An early study of the
evolution of evaporating black holes in inflationary cosmology is in
\cite{mallet}. 

We begin in Sec.~II to present the relevant solutions to
which the method will be applied. This  will actually be done in
Sec.~III, where we shall also display the ``generalized first law''
and the  conclusions to be drawn from it.

\section{Dynamical black holes, and a fake one}

We consider first spherically symmetric spacetimes which outside the
horizon (if there is one) are described by a metric of the form
\be
ds^2=-e^{2\Psi(r,v)}A(r,v)dv^2+2 e^{\Phi(r,v)}dvdr+
r^2dS^2_{D-2}\,.
\label{efbard}
\ee
where the coordinate $r$ is the areal radius commonly used in relation
to spherical symmetry and $v$ is intended to be an advanced null
coordinate. In an asymptotically flat context one can always write (we
use geometrized units in which the Newton constant $G=1$) 
$A(r,v)=1-2m(r,v)/r^{D-3}$. 
This metric was first proposed by Vaidya
\cite{vaidya} in $D=4$ dimensions, and studied in an interesting paper
during the classical era of black hole physics by Lindquist et al
\cite{lind}. It has been generalized to Einstein-Maxwell systems and
de Sitter space by Bonnor-Vaidya and Mallet, respectively \cite{bvm}. 
In the special case $\Phi(r,v)=\Psi(r,v)$  it was then
extensively used by Bardeen \cite{bardeen}  
and York \cite{york} in their semi-classical analysis of the back 
reaction problem, 
the former to establish the stability of the event horizon in the
geometry modified by Hawking radiation back reaction (contrary to a
claim of F.~Tipler), the latter in an attempt to explain dynamically
the origin of the entropy of the black holes. We shall call it the
Vaidya-Bardeen metric. If one wishes the metric can also be
written in double-null form. In the $(v,r)$-plane one can introduce
null coordinates $x^{\pm}$  such that the dynamical Vaidya-Bardeen BHs
may be written as  
\be
ds^2=-2f(x^+,x^-)dx^+dx^-+r^2(x^+,x^-)dS^2_{D-2}\,,
\label{nf}
\ee
for some differentiable function $f$. The remaining angular 
coordinates contained in $dS^2_{D-2} $ do not play any essential 
role. In the following we shall use both forms of the metric,
depending on computational convenience. The field equations of
the Vaidya-Bardeen metric in $D=4$ dimensions are of interest. They
read   
\be\label{vbe}
\frac{\partial m}{\partial v}=4\pi r^2T^r_{\;v}, \quad \frac{\partial
  m}{\partial r}=-4\pi r^2T^v_{\;v}, \quad 
 \frac{\partial\Psi}{\partial r}=4\pi re^{\Psi}T^v_{\;r}
\ee
The second example we are interested in is the McVittie solution
\cite{mcv} for a point mass in a Friedmann-Robertson-Walker flat
cosmology. In D-dimensional  spacetime in isotropic spatial
coordinates it is given by \cite{gao} 
\be
ds^2=-A(\rho,t)dt^2+B(\rho,t)\at d\rho^2+\rho^2dS_{D-2}^2 \ct\,
\ee
with
\be
A(\rho,t)=\aq \frac{1-\at \frac{m}{a(t)\rho}\ct^{D-3}}{1+
\at \frac{m}{a(t)\rho}\ct^{D-3}} \cq^2\,,\qquad
 B(\rho,t)=a(t)^2 \aq 1-\at \frac{m}{a(t)\rho}\ct^{D-3}\cq^{2/(D-3)}\,.
\ee
When the mass parameter $m=0$, it reduces to a spatially flat FRW
solution with scale factor $a(t)$; when $a(t)=1$ it reduces to the
Schwarzschild metric with mass $m$.  
In four dimensions this solution has had a strong impact on the
general problem of matching the Schwarzschild solution with cosmology,
a problem faced also by Einstein and Dirac. 
Besides McVittie, it has been extensively studied by Nolan 
in a series of papers \cite{nolan}. To put the metric in the general
form of Kodama theory, we use what may be called the Nolan gauge, in
which  the metric reads 
\be\label{nolan}
ds^2=-\at A_s-H^2(t)r^2 \ct dt^2+A_s^{-1}dr^2-2A_s^{-1/2}
H(t)r\,drdt +r^2dS^2_{D-2}
\ee
where $H(t)=\dot{a}/a$ is the Hubble parameter and, for example, in the
charged 4-dimensional case, $A_s=1-2m/r+q^2/r^2$ and in $D$ dimension 
$A_s=1-2M/r^{D-3}+Q^2/r^{2D-6}$. 
 In passing to the Nolan gauge a choice of sign in the cross term
$drdt$ has been done, corresponding to an expanding universe; the
transformation $H(t)\to-H(t)$ changes this into a contracting one. 
In the following we shall consider $D=4$ and $q=0$; then the
Einstein-Friedmann equations  read
\be\label{eeqs}
3H^2=8\pi\rho\,, \qquad 2A_s^{-1/2}\dot H(t)+3H^2=-8\pi p\,.
\ee
It follows that $A_s=0$, or $r=2m$, is a curvature singularity. In
fact, it plays the role that 
$r=0$ has in FRW models,  namely it is a big bang
singularity.  
When $H=0$ one has the Schwarzschild solution. Note how the term
$H^2r^2$ in the metric strongly resembles a varying cosmological
constant; in fact for $H$ a 
constant, it reduces to the Schwarzschild-de Sitter solution in Painlev\'e
coordinates. As we will see, the McVittie solution possesses in general 
both apparent and trapping horizons, and the spacetime is
dynamical. However, it is really not a dynamical black hole in the
sense we used it above, since the mass parameter is strictly constant:
for this reason we called it a fake dynamical BH. This observation
prompts one immediately for an obvious extension of the solution: to
replace the mass parameter by a function of time and radius, but this
will not be pursued here.  

The solutions being given, one may ask whether they have anything to
do with black holes. Unfortunately it is not entirely clear what
should be considered a black hole in a dynamical regime. However, 
some kind of horizon must be
present, and moreover, the evolution should be sufficiently slow to
permit a comparison with the more familiar stationary case.   
In order for a horizon to exist some metric components must have a
zero somewhere. To clarify the 
issue we have to introduce few definitions, so at this point we have
to refer the reader to references \cite{Hayward:1993wb}, 
\cite{Hayward:1997jp}, \cite{Hayward:1994bu}, \cite{hay} and
\cite{abbey} for more 
 details. The expansions $\theta_{\pm} $ of the null geodesic
congruences orthogonal to a closed surface $S$ with codimension two
and measure $\mu(S)=\Omega_{D-2}r^{D-2}$, are defined by 
$\theta_{\pm}=\mu(S)^{-1}\partial_{\pm}\mu
(S)=(D-2)r^{-1}\partial_{\pm}r$. 
A marginal surface $S$ is a $(D-2)$-dimensional spacelike surface with
vanishing expansion, and a dynamical horizon is a hypersurface which is
foliated by marginal surfaces. It is possible to show that the area
of such marginal surfaces is non decreasing if the energy inflow is
non negative. A future dynamical
horizon, say $H^+$, is the hypersurface implicitly determined by the
condition $\theta_+=0$\footnote{There are dual definitions involving
$\theta_-$ and some interchanging of ``future'' with ``past'', see 
the references for details.}. It will be called a trapping horizon if  
$\theta_+$ is strictly decreasing on crossing the horizon from
the outside (so that $\partial_-\theta_+<0$ on $H^+$), a
condition which insures the non vanishing of the surface gravity 
(indeed it is violated for extremal black holes and naked
singularities).  
We just mentioned the surface gravity; a  geometrical definition of this
quantity for a trapping horizon is \cite{Hayward:1997jp}, 
$\kappa=(D-2)^{-2}g^{+ -}\aq (D-2)\partial_- \theta_++\theta_+\theta_- 
\cq_{|\theta_+=0}$. 
We stress that this quantity is not the same surface gravity that was
defined few years before this new proposal, although it fitted equally
well to a generalized first law. Later
on we will show that $\kappa$ fixes the expansion of the
metric near the horizon along a future null direction.

These definitions look somewhat artificial, but in fact they are very
natural and connected directly with what 
is known for the stationary black holes. To see this one notes,
following Kodama \cite{Kodama:1979vn}, that any metric like
\eqref{efbard} or \eqref{nf}, admits a
unique (up to normalization) vector field $K^a$ such that $K^aG_{ab}$
is divergence free, where $G_{ab}$ is the Einstein tensor; for
instance, using the double-null form, one finds
$K=-g^{+-}(\partial_+r\partial_--\partial_-r\partial_+)$, 
while using \eqref{efbard} one has $K=e^{-\Phi}\partial_v$. In 
any case the defining property of $K$ show that it is a natural
generalization of the time translation Killing field of a static
black hole. Moreover, by Einstein equations $K_aT^{ab}$ will be
conserved so for such metrics there exists a natural (Kodama wrote
``preferable'') localizable energy flux and its conservation law. Now
consider the expression $K^a\nabla_{[b}K_{a]}$: it is not hard to see  
that on $H^+$ it is proportional to $K_b$. The proportionality
factor, a function in fact, is the surface gravity: 
$K^a\nabla_{[b}K_{a]}=-\kappa K_b$. 
For a Killing vector field $\nabla_{b}K_{a}$ is anti-symmetric so the
definition reduces to the usual one. 
The study of black holes requires  also a notion of energy; the natural
choice would be to use the charge associated to Kodama conservation
law, but this turns out to be the Misner-Sharp energy, which for a
sphere with areal radius $r$ is the same as the Hawking
mass\cite{hawk} (here $D=4$),   
$E=2^{-1}r(1-2^{-1}r^2g^{+-}\theta_+\theta_-)$.   
If this definition is found not particularly illuminating, then one
can use the metric \eqref{efbard} to obtain the equivalent expression
$g^{\mu\nu}\partial_{\mu}r\partial_{\nu}r=1-2E/r$, 
where $g_{\mu\nu}$ is the reduced metric in the plane normal to the
sphere of symmetry (the plane $(v,r)$). 
In this form it is clearly a generalization of the Schwarzschild
mass. As we said, $E$ is just the charge associated to Kodama conservation
law; as showed by Hayward \cite{Hayward:1994bu}, in vacuo $E$ is
also the Schwarzschild energy, at null infinity it is the Bondi-Sachs
energy and at spatial infinity it reduces to the ADM mass.

Let us apply this general theory to the two classes of dynamical BH we
have considered. Using Eq.~\eqref{efbard}, we have 
$\theta_+=(D-2)e^{2\Psi-\Phi} A(r,v)/2r\,.$ The condition $\theta_+=0$
leads to $A(r_H,v)=0$,  
which defines a curve $r_H=r_H(v)$ giving the location of the 
apparent horizon; writing the solution in the Vaidya-Bardeen
form, that is with $A(r,v)=1-2m(r,v)/r$, the Misner-Sharp energy of the
black hole is $E=m(r_H(v),v)$, and the horizon will be trapping if
$m^{'}(r_H,v)<1/2$, a prime denoting the radial derivative.
The  surface gravity associated with the Vaidya-Bardeen  dynamical
horizon is
\be\label{vbk}
\kappa(v)=\frac{A^{'}(r,v)}{2}|_{r=r_H}=\frac{m(r_H,v)}{r_H^2}-
\frac{m^{'}(r_H,v)}{r_H}=\frac{1}{2r_H}-\frac{m^{'}(r_H,v)}{r_H}\,.
\ee
We see the meaning of the trapping condition: it ensures the
positivity of the surface gravity. In the case of McVittie BHs, we
obtain $\theta_{\pm}=\pm(D-2)(\sqrt A_s\mp H r)/2rf_{\pm}$, 
where the functions $f_{\pm}$ determine null coordinates $x^{\pm}$
such that $dx^{\pm}=f_{\pm} \aq \at \sqrt A_s\pm H r \ct dt\pm
A_s^{-1/2}dr\cq$.  One may compute from this the dual derivative
fields $\partial_{\pm}$.  
Thus, the future dynamical horizon  defined by $\theta_+=0$, has a
radius which is a root of the equation $\sqrt A_s=H r_H \,,$
which in turn implies $A_s=H^2r^2_H$. Hence the horizon radius is a
function of time. The Misner-Sharp mass and the
related surface gravity are
\be
E=m+\frac{1}{2}\,H(t)^2r_H^3\,, \quad
\kappa(t)=\frac{1}{2}\aq A^{'}_s(r_H)\cq -H^2 r_H -\frac{ \dot H}{2 H}=
\frac{m}{r_H^2}-H^2 r_H -\frac{ \dot H}{2 H}\,.
\label{mvsg}
\ee
Note that $E=r_H/2$. In the static cases everything agrees with the
standard  results. The 
 surface gravity has an interesting expression in terms of the sources
of Einstein equations and the Misner-Sharp mass. Let $T_{(2D)}$ be the
reduced trace of the stress tensor in the space normal to the sphere
of symmetry, evaluated on the horizon $H^+$. For the Vaidya-Bardeen
metric it is, by Einstein's equations \eqref{vbe},
\[
T_{(2D)}=T^v_{\;v}+T^r_{\;r}=-\frac{1}{2\pi r_H}\,\frac{\partial
  m}{\partial r}_{|r=r_H}
\]
For the McVittie's solution, this time by Fredmann's equations
\eqref{eeqs} one has  
\[
T_{(2D)}=-\rho+p=-\frac{1}{4\pi}\at 3H^2+\frac{\dot{H}}{Hr_H}\ct
\]  
We have then the formula, $\kappa=r_H^{-2}E+2\pi r_HT_{(2D)}$. 
It is worth mentioning the pure FRW case, i.e. $A_s=1$, for which
$\kappa(t)= -\at H(t)+\dot H/2 H\ct\,.$
We feel that these expressions for the surface gravity are non trivial
and display deep connections with the emission process. Indeed it is
the non vanishing of $\kappa$ that is connected with the imaginary
part of the action of a massless particle, as we are going to show in
the next section. 

\section{Tunneling within the Hamilton-Jacobi method, and the
  conclusions} 

The essential property of the tunneling method is that the action
$I$ of an outgoing massless particle emitted from the horizon has an
imaginary part which for stationary black holes is $\Im
I=\pi\kappa^{-1}E$, where $E$ is the Killing energy and $\kappa$ the 
horizon surface gravity. The imaginary part is obtained by means of
Feynman $i\epsilon$-prescription, as explained in
\cite{parikh,angh}. As a result the particle production rate reads  
$\Gamma= \exp(-2 \Im I)=\exp(-2\pi\kappa^{-1}E)\,.$
One then recognizes the Boltzmann factor, from which one deduces the
well-known temperature $T_H=\kappa/2\pi$.
But more than this, an explicit expression for
$\kappa$ is actually obtained in terms of radial derivatives of the
metric on the horizon.  
Let us consider now the case of a dynamical black hole in the double-null  
form. We have for a massless particle along a radial geodesic the
Hamilton-Jacobi equation $\partial_+I \partial_-I=0\,.$ 
Since the particle is outgoing $\partial_- I$ is not vanishing, and we
arrive at the simpler condition  $\partial_+ I=0$. 
First, let us apply this condition to the Vaidya-Bardeen BH.  One
has then
\be\label{hjv}
2e^{-\Psi(r,v)}\partial_v I+A(r,v)\,\partial_r I=0\,.
\ee
Since the particle will move along a future null geodesic, to pick
the imaginary part  we expand the metric along a future null direction
starting from an arbitrary event $(r_H(v_0),v_0)$ on the horizon,
i.e. $A(r_H(v_0),v_0)=0$.  Thus, shortening $r_H(v_0)=r_0$,
we have 
$A(r,v)=\partial_r A(r_0,v_0)\Delta r+\partial_vA(r_0,v_0)\Delta
v\dots=2\kappa(v_0)(r-r_0)+\dots$,  
since along a null direction at the horizon $\Delta v=0$, according to 
 the metric \eqref{efbard}; here $\kappa(v_0)$ is the surface gravity, 
Eq.~\eqref{vbk}. 
From \eqref{hjv} and the expansion, $\partial_rI$ has a simple pole at
the event $(r_0,v_0)$; as a consequence
\be
\Im I=\Im \int  \partial_r I dr=-\Im  \int dr
\frac{2e^{-\Psi(r,v)}\partial_v I}{A^{'}(r_0,v_0)(r-r_0-i0)} 
=\frac{\pi \omega(v_0) }{\kappa(v_0)}\,. 
\ee
where  $\omega(v_0)= e^{-\Psi(r_0,v_0)}\partial_vI $, is to be
identified with  the energy of the particle at the time $v_0$. Note that
the Vaidya-Bardeen metric has a sort of gauge invariance due to conformal
reparametrizations of the null coordinate $v$: the map $v\to\tilde
v(v)$, $\Psi(v,r)\to\tilde\Psi(\tilde v,r)+\ln(\partial\tilde v/\partial
v)$ leaves the metric invariant, and the energy is gauge invariant
too. Thus we see that it is the Hayward-Kodama surface gravity that is
relevant to the process of particles emission. The emission
probability, $\Gamma= \exp(-2\pi\omega(v)/\kappa(v))$, has the form of
a Boltzmann factor, suggesting a locally thermal spectrum.
For the McVittie BH, the situation is similar. In fact, the
condition $\partial_+I=0$ becomes 
$\partial_rI=-F(r,t)^{-1}\partial_t I\,,$
where $F(r,t)=\sqrt{A_s(r)}(\sqrt{A_s(r)}-rH(t))$. As before, we pick
the imaginary part by expanding this function at the  horizon  
along a future null direction, using the fact that for two
neighbouring events on a null direction in the metric \eqref{nolan},
one has $t-t_0=(2H_0^2r_0^2)^{-1}(r-r_0)$, where $H_0=H(t_0)$. We find
the result 
\be\label{exp1}
F(r,t)=\at\frac{1}{2}\,A^{'}_s(r_0)-r_0 H_0^2-\frac{\dot H_0}{2H_0}\ct
(r-r_0)=\kappa(t_0)(r-r_0)\dots 
\ee
where this time $r_0=r_H(t_0)$.
From this equation we see that $\partial_rI$ has a simple pole at the
horizon; hence, making use again of Feynman $i\epsilon$-prescription,
one finds $\Im I=\pi\kappa(t_0)^{-1}\omega(t_0)$, where
$\omega(t)=\partial_tI$ is again the energy at time $t$,   
in complete agreement with the geometric evaluation of the previous
section. Obviously, if $\kappa$ vanishes on the horizon there is no
simple pole and the black hole should be stable\footnote{However,
  charged  extremal black holes can radiate \cite{Vanzo:1995bh}.}. The
kind of instability producing the Hawking flux for 
stationary black holes evidently persists in the dynamical arena, and
so long 
as the evolution is sufficiently slow the black hole seems ``to
evaporate thermally'' (paraphrasing \cite{bardeen}). Note that the
imaginary part, that is the instability, is attached to the horizon
all the time, confirming the Fredenhagen-Haag suggestion quoted in the
introduction. It is worth
mentioning the role of $\kappa$ in the analogue of the first law for
dynamical black holes (contributions to this problem for Vaidya black
holes were given in \cite{cinesi}). Using the formulas of the
projected stress tensor $T_{(2D)}$ given above, and the expression of
the Misner-Sharp energy, one obtains the differential law 
\be\label{fl}
dE=\frac{\kappa\, dA_H}{8\pi}-\frac{T_{(2D)}}{2}\,dV_H
\ee
provided all quantities were computed on the horizon.
Here $A_H=4\pi r_H^2$ is the horizon area and $V_H=4\pi r_H^3/3$ is a
formal horizon volume. If one interprets the ``d'' operator as
a derivative along the future null direction one gets Hayward's form
of the first law. But one can also interpret the differential
operation more abstractly, as referring to an ensemble. Indeed, to
obtain Eq.~\eqref{fl} it is not necessary to specify the meaning of the
``d''. It is to be noted that the same law can be proved with other,
inequivalent definitions of the surface gravity, even maintaining the
same meaning of the energy. Thus it was not a trivial problem to identify
the correct one: the tunneling method has made the choice.


\begin{thebibliography}{99}

\bibitem{Fredenhagen:1989kr}
  K.~Fredenhagen and R.~Haag,
  %``ON THE DERIVATION OF HAWKING RADIATION ASSOCIATED WITH THE FORMATION OF A
  %BLACK HOLE,''
  Commun.\ Math.\ Phys.\  {\bf 127}, 273  (1990).
  %%CITATION = CMPHA,127,273;%%

\bibitem{vaidya}
P.~C.~Vaidya, Proc.~Indian Acad.~Sci.~{\bf A33}, 264 (1951); 
P.~C.~Vaidya, Nature {\bf 171}, 260 (1953); 
 V.V. Narlikar and P.C. Vaidya,
   Nature {\bf 159}, 642 (1947).

\bibitem{bardeen} J.M. Bardeen, Phys.~Rev.~Letters {\bf 46},  
382 (1981).
\bibitem{york} J. W. York, Phys.~Rev.~{\bf D 28}, 2929 (1983).

\bibitem{mcv}G. C. McVittie , Mon.~Not.~R.~Astronomic Soc.~{\bf 93},  
325 (1933).

\bibitem{parikh}
M.~K.~Parikh and F.~Wilczek,
%``Hawking Radiation as Tunneling,''
Phys.\ Rev.\ Lett.\  {\bf 85}, 5042  (2000).

\bibitem{svmp}
S.~Shankaranarayanan, T.~Padmanabhan and K.~Srinivasan,
%``Hawking Radiation in Different Coordinate Settings: Complex Paths
%Approach'',  
Class.\ Quant.\ Grav.\  {\bf 19}, 2671 (2002); 
E.~C.~Vagenas,
%``Generalization of the KKW Analysis for Black Hole Radiation,''
 Phys.\ Lett.\  {\bf B 559}, 65  (2003); 
A.~J.~M.~Medved,
%``Radiation Via Tunneling from a de Sitter Cosmological Horizon,''
Phys.\ Rev.\ {\bf D66}, 124009  (2002); 
 T.~Padmanabhan,
%``Entropy of horizons, complex paths and quantum tunneling,''
  Mod.\ Phys.\ Lett.\ A {\bf 19}, 2637 (2004).


\bibitem{others}

  K.~Srinivasan and
  T.~Padmanabhan , %Particle Production and Complex Path Analysis, 
  Phys.~Rev.~{\bf D 60}, 24007 (1999);
  S.~Hemming and E.~Keski-Vakkuri,
  %``Hawking radiation from AdS black holes,''
  Phys.\ Rev.\  D {\bf 64}, 044006 (2001)
  [arXiv:gr-qc/0005115];
A.~J.~M.~Medved and E.~C.~Vagenas,
  %``On Hawking radiation as tunneling with back-reaction,''
  Mod.\ Phys.\ Lett.\  A {\bf 20}, 2449 (2005); 
M.~Arzano, A.~J.~M.~Medved and E.~C.~Vagenas,
  %``Hawking radiation as tunneling through the quantum horizon,''
  JHEP {\bf 0509}, 037 (2005); 
Q.~Q.~Jiang, H.~L.~Li, S.~Z.~Yang and D.~Y.~Chen,
  %``Hawking tunneling radiation of black holes in de Sitter and anti-de Sitter
  %spacetimes,''
Mod.\ Phys.\ Lett.\  A {\bf 22}, 891 (2007).

\bibitem{anm}
E.~T.~Akhmedov, V.~Akhmedova and D.~Singleton,
  %``Hawking temperature in the tunneling picture,''
  Phys.\ Lett.\  B {\bf 642}, 124 (2006); T.~K.~Nakamura, arXiv:
  0706.2916 [hep-th]; P.~Mitra, Phys.Lett. {\bf B648} 240 (2007).


\bibitem{Hayward:1993wb}
  S.~A.~Hayward,
  %``General laws of black hole dynamics,''
  Phys.\ Rev.\  D {\bf 49}, 6467 (1994).

 
\bibitem{angh}
%HAWKING RADIATION AS TUNNELING FOR EXTREMAL AND ROTATING BLACK HOLES.
M. Angheben, M. Nadalini, L. Vanzo and S. Zerbini,
JHEP {\bf 0505}, 014 (2005); 
% HAWKING RADIATION AS TUNNELING: THE D DIMENSIONAL ROTATING CASE
      M. Nadalini, L. Vanzo and S. Zerbini, 
      J. Physics A Math. Gen. {\bf 39}, 6601 (2006). 
\bibitem{mann}
B. Kerner and R. B. Mann, Phys. Rev. {\bf D 76}, 104010 (2006).

\bibitem{Fodor:1996rf}
  G.~Fodor, K.~Nakamura, Y.~Oshiro and A.~Tomimatsu,
  %``Surface gravity in dynamical spherically symmetric spacetimes,''
  Phys.\ Rev.\  D {\bf 54}, 3882 (1996).

\bibitem{Hayward:1997jp}
 S.~A.~Hayward,
  %``Unified first law of black-hole dynamics and relativistic
 %thermodynamics,'' 
  Class.\ Quant.\ Grav.\  {\bf 15}, 3147 (1998).


\bibitem{Kodama:1979vn} 
  H.~Kodama,
  %``Conserved Energy Flux For The Spherically Symmetric System And The Back
  %Reaction Problem In The Black Hole Evaporation,''
  Prog.\ Theor.\ Phys.\  {\bf 63}, 1217 (1980).

\bibitem{isohor}
Abhay Ashtekar and Badri Krishnan,
"Isolated and Dynamical Horizons and Their Applications",
Living Rev. Relativity 7,  (2004),  10. URL:
http://www.livingreviews.org/lrr-2004-10; A.~Ashtekar, A.~Corichi and
K.~Krasnov, %``Isolated horizons: The classical phase space,''
  Adv.\ Theor.\ Math.\ Phys.\  {\bf 3}, 419 (2000);  A.~Ashtekar,
  S.~Fairhurst and B.~Krishnan, 
  %``Isolated horizons: Hamiltonian evolution and the first law,''
  Phys.\ Rev.\  D {\bf 62}, 104025 (2000);  A.~Ashtekar, C.~Beetle,
  O.~Dreyer, S.~Fairhurst, B.~Krishnan, J.~Lewandowski and
  J.~Wisniewski, 
  %``Isolated horizons and their applications,''
  Phys.\ Rev.\ Lett.\  {\bf 85}, 3564 (2000).

\bibitem{gaowu}
Xiaoning Wu, Sijie Gao, Phys.~Rev.~{\bf D 75}, 044027-1 (2007).

\bibitem{mallet}
R.~L.~Mallet, Phys.~Rev.~{\bf D 33}, 2201 (1986).

\bibitem{lind}
R.~W.~Lindquist, R.~A.~Schwartz, C.~W.~Misner, Phys.~Rev.~{\bf 137},
1364 (1965).



\bibitem{bvm}
W.~B.~Bonnor, P.~C.~Vaidya, Gen.~Rel.~Grav.~{\bf 1}, 127, (1970); 
R.~L.~Mallet, Phys.~Rev.~{\bf D31}, 416 (1985).

\bibitem{gao}
  C.J.~Gao,
    Class.~Quant.~Grav.~{\bf 21}, 4805 (2004).

\bibitem{nolan}
  B.~C.~Nolan,
  %``A point mass in an isotropic universe: Existence, uniqueness and basic
  %properties,''
  Phys.\ Rev.\  D {\bf 58}, 064006 (1998); 
B.~C.~Nolan,
  %``A Point mass in an isotropic universe: II Global properties,''
  Class.\ Quant.\ Grav.\  {\bf 16}, 1227 (1999); B.~C.~Nolan,
  %``A point mass in an isotropic universe: III. The region $R\leq 2m$,''
  Class.\ Quant.\ Grav.\  {\bf 16}, 3183 (1999).


\bibitem{Hayward:1994bu}
  S.~A.~Hayward,
  %``Gravitational energy in spherical symmetry,''
  Phys.\ Rev.\  D {\bf 53}, 1938 (1996).

  %%CITATION = PHRVA,D53,1938;%%
\bibitem{hay}
 S. A. Hayward,
    Phys.\ Rev. Letters \ D {\bf 93}, 251101 (2004).
\bibitem{abbey}
A. Ashtekar and B. Krishnan, 
 Phys.\ Rev. Letters \ D {\bf 89}, 261101 (2002); 
A. Ashtekar and B. Krishnan, Living Rev. Rel. {\bf 7}, 10 (2004).

%\bibitem{Misner:1964je}

\bibitem{hawk}
  C.~W.~Misner and D.~H.~Sharp,
  %``Relativistic equations for adiabatic, spherically symmetric gravitational
  %collapse,''
  Phys.\ Rev.\  {\bf 136}, B571 (1964); 
  %%CITATION = PHRVA,136,B571;%%
S.~W.~Hawking, J.~Math.~Phys.~(N.Y.) {\bf 9},  598 (1968).

\bibitem{Vanzo:1995bh}
  L.~Vanzo,
  %``Radiation from the extremal black holes,''
  Phys.\ Rev.\  D {\bf 55}, 2192 (1997).
  %%CITATION = PHRVA,D55,2192;%%

\bibitem{cinesi}
Li Xiang, Zhao Zheng, Phys.~Rev.~{\bf D 62}, 104001-1 (2000); 
Xiaoning Wu, Sijie Gao, Phys.~Rev.~{\bf D75}, 044027 (2007);  Ren
Ji-Rong, Li Ran, arXiv:0705.4339 [gr-qc].


%\bibitem{mitra}
%P. Mitra, Phys.Lett. {\bf B648} 240 (2007).
%B. Chatterjee, A. Ghosh, P. Mitra,
%Tunnelling from black holes in the Hamilton Jacobi approach

 
\end{thebibliography}
\end{document}